\documentclass{ar-1col}
\usepackage{cite}
\usepackage{amsmath}
\usepackage{amssymb}
\usepackage{bm}
\usepackage{tikz}

\usepackage{url}
\usepackage{soul,upgreek}
\jname{Annu. Rev. Nucl. Part. Sci.} \jvol{68} \jyear{2018}
\doi{10.1146/annurev-nucl-101917-021019}

\newcommand{\pt}{p_{\rm T}}

\newcommand{\bmx}{\bm{x}}
\newcommand{\bmz}{\bm{z}}
\newcommand{\bmy}{\bm{y}}
\newcommand{\bmX}{\bm{X}}
\newcommand{\bmY}{\bm{Y}}

\begin{document}

\markboth{Guest $\bullet$ Cranmer $\bullet$ Whiteson}{Deep Learning and Its Application to LHC
Physics}

\title{Deep Learning and Its Application to LHC Physics}

\author{Dan Guest,$^1$ Kyle Cranmer,$^2$ and Daniel Whiteson$^1$
\affil{$^1$Department of Physics and Astronomy, University of California, Irvine, California 92697, USA}
\affil{$^2$Physics Department, New York University, New York, NY 10003, USA}}

\begin{abstract}
Machine learning has played an important role in the analysis of high-energy physics data for decades. The emergence of deep learning in 2012 allowed for machine learning tools which could adeptly handle higher-dimensional and more complex problems than previously feasible. This review is aimed at the reader who is familiar with high energy physics but not machine learning.  The connections between machine learning and high energy physics data analysis are explored, followed by an introduction to the core concepts of neural networks, examples of the key results demonstrating the power of deep learning for analysis of LHC data, and discussion of future prospects and concerns.
\end{abstract}


\maketitle

\tableofcontents

\section{INTRODUCTION}

The physics program of the Large Hadron Collider (LHC) has the
potential to address many of the most fundamental questions in
modern physics: the nature of mass, the dimensionality of space,
the unification of the fundamental forces, the particle nature of
dark matter, and the fine-tuning of the Standard Model.  The
importance of these questions and the scale of the experimental
program needed to probe them demand that we do our utmost to
extract the relevant information from the collected data.

The data collected by high-energy physics (HEP) experiments are
complex and high dimensional. Traditional data analysis techniques
in HEP use a sequence of boolean decisions followed by statistical
analysis on the selected data. Typically, both the individual
decisions and the subsequent statistical analysis are based on the
distribution of a single observed quantity motivated by physics
considerations, which is not easily extended to higher dimensions.

For several decades, particle physicists have sought to improve
the power of their analyses by employing algorithms that  utilize
multiple variables simultaneously. Within HEP this approach is
often referred to as {multivariate analysis} (MVA);
however, outside of physics these techniques would be considered
examples of {machine learning}. Physicists have used a wide
variety of machine learning techniques, including artificial neural
networks, kernel density estimation, support vector machines,
genetic algorithms, random forests, and boosted decision trees.
For several years, the status quo of machine learning in HEP was
to use boosted decision trees implemented in the software package
TMVA~\cite{Hocker:2007ht}. These tools provided an important boost
for many data analysis tasks, but their capabilities were
understood to be limited; they often failed to match the
performance of physicist-engineered solutions, especially when the
dimensionality of the data grew large.

The emergence of {deep learning} began around 2012, when a convergence of techniques enabled training of very large neural networks that greatly outperformed the previous state of the art~\cite{krizhevsky2012imagenet, imagenet-challenge,lecun2015deep,schmidhuber2015deep}. These new tools could adeptly handle higher-dimensional and more complex problems than previously feasible. In the intervening years there has been an explosion in deep learning research moving beyond the application to image classification into natural languages, self driving cars, and many areas of science.  

This review is aimed at the reader who is familiar with data analysis for high energy physics but less familiarity with machine learning.  The remainder of this section explains the importance of machine learning to high-energy physics data analysis and describes the basics of neural networks in order to clearly define the concept of deep learning.  Section 2 reviews many of the key applications of deep learning to open challenging problems in LHC data analysis, including several breakthroughs in tasks which were previously thought to be intractable. Section 3 discusses the direction of current work and potential concerns regarding the application of deep learning. The final section discusses several possible future directions and prospects.

\subsection{Why Is Machine Learning Relevant for Physics?}

The data collected by the LHC experiments are vast in both the number of collisions and in the complexity of each collision. The colliding beams at the LHC are grouped into bunches of protons, which cross with a frequency of  $\sim$40 MHz. Each collion has the potential to produce a large number of particles, and the LHC detectors have $\mathcal{O}(10^8)$ sensors used to record these particles.  These high data rates are necessary because collisions which produce interesting products are very rare.

As a result of the quantum-mechanical nature of the collisions and
the interaction of their products with LHC detectors, the
observations resulting from a particular interaction are
fundamentally probabilistic.  Therefore, the approach to data
analysis and the conclusions drawn from the data must be framed in
statistical terms, which includes not only low-level tasks such as
particle identification and reconstruction of the particles' energy
and momentum, but also  high-level tasks such as searches for new
particles and measurements.  In classical statistics, tasks such
as classification, hypothesis testing, regression, and
goodness-of-fit testing are based a statistical model $p(\bmx |
\theta)$ describing the probability of observing $\bmx$ given the
parameters of a theory $\theta$.

The high dimensionality and large volume of LHC data pose a
problem because the  statistical model $p(\bmx  | \theta)$ over
the high-dimensional space of their experimental data is not
known explicitly in terms of an equation that can be evaulated. Instead, one typically has access to  large samples of
simulated data that was generated by stochastic
simulation programs that model the physics of particle
interactions on various scales. If the data were fairly
low dimensional ($d<5$), the problem of estimating the unknown
statistical from the simulated samples would not be difficult.
Histograms or kernel-based density estimates provide reasonable
estimates in low-dimensional spaces.   These fairly na\"ive
strategies, however, suffer from the {curse of
dimensionality}. In a single dimension,  $N$ samples
may be required to describe the source probability density function. In $d$ dimensions, the number of
samples required grows to the power of the data's dimensionality:
$\mathcal{O}(N^d)$. The consequence is that any dimensionality
greater than five or ten requires impractical or impossible
computational resources, regardless of the speed of the sample
generator.

Traditionally, HEP physicists have approached this problem by
reducing the dimensionality of the data through a series of steps
that operate both on  individual collision events and on
collections of events. For an individual event,
{reconstruction} algorithms process the raw sensor data
into low-level objects such as calorimeter clusters and tracks.
From these low-level components, the algorithms attempt to
estimate the energy,  momentum, and identity of individual
particles. From these reconstructed objects, event-level summaries
are constructed. Event selection algorithms then select subsets of
the collision data for further analysis on the basis of the information
associated to individual events. Traditionally, the reconstruction
and event selection operations are based
on specific, engineered features in the data. For instance, the
identification of an electron and a photon is based on specific
features that summarize the shape of the shower in the
electromagnetic calorimeter and discriminate from the energy
deposits left by charged and neutral hadrons. The cumulative
product of these steps reduces the dimensionality of the problem to
a number small enough to allow the missing statistical model $p(\bmx
| \theta)$ to be estimated using samples generated by simulation
tools.

While the traditional approaches to reconstruction and event
selection have worked fairly well, there is no guarantee that they
are optimal. Given the complex nature of the data and the subtle
signatures of potential new physics, it is reasonable to suspect
that there may be a significant performance gap between
traditional approaches and the optimal one.
A central role of machine learning in LHC physics is to improve
this data reduction, reducing the relevant information
contained in the low-level, high-dimensional data into a higher-level and smaller-dimensional
space.

\subsection{The Role of Simulators}

Physicists often refer to the set of simulation tools such as
\textsc{pythia}~\cite{Sjostrand:2006za},
\textsc{herwig}~\cite{Bahr:2008pv},
\textsc{MadGraph}~\cite{Alwall:2014hca},
\textsc{Sherpa}~\cite{Gleisberg:2008ta}, and
\textsc{Geant}~\cite{Agostinelli:2002hh} as {Monte Carlo}
tools, since these simulations are probabilistic and rely heavily
on Monte Carlo sampling techniques. The simulators capture the
relevant physics on a hierarchy of scales starting with the
microscopic interactions within a proton--proton collision and
ending with the interaction of particles in the enormous LHC
detectors.\footnote{In the language of statistics and machine learning, the
full simulation chain would be considered a  {generative
model} for the data as they can be used to generate synthetic data
of the same complexity and format as the actual collision data.}

We can think of a simulated data set $\{\bmx_i\}_{i=1}^N$ as being $N$
independent and identically distributed samples from some
underlying distribution $p(\bmx |\theta)$, where $\theta$
corresponds to the settings of the simulator. Moreover, we know
the settings of the simulator used, which means that the generated data
automatically come with ground-truth labels. For instance, we can
generate samples of interactions involving a Higgs boson for any
desired mass value.

In this framing, the goal of simulation is to approximate the probability $p(\bmx | \theta)$ by sampling from an enormous space of unobserved, or {latent}, processes:
$p(\bmx |\theta) = \int p(\bmx,\bmz | \theta) \mathrm{d}\bmz$. A fixed value of $\bmz$ specifies
everything about the simulated events, from the momentum of the
initial particles created in the hard scattering to the detailed
interactions in the detector. Physicists often refer to $\bmz$ as {Monte Carlo truth}. Most
reconstruction algorithms can be regarded as estimates of some
components of $\bmz$ (particle type, momentum, energy, etc.) given
the observed data $\bmx$. Here simulation fulfills a second experimental need: in addition to an estimate of $p(\bmx | \theta)$, the simulation provides a dataset $\{\bmx_i,\bmz_i\}_{i=1}^N$ which allows physicists to study reconstruction algorithms directly.

\subsection{Core Concepts in Machine Learning}

Fortunately, many of the tasks encountered in high energy physics can be naturally reformulated as machine learning problems. Typically, the problems are formulated in terms of a search for some function $f:\bmX\to \bmY$, from the space of the observed data $\bmX$  to a low-dimensional space of a desired target label $\bmY$, which optimizes some metric of our choosing. This metric is called a {loss function} and written as $L(\bmy,f(\bmx ))$.

Ideally, a learning algorithm would find the function that optimizes $L$ over all possible values of $(\bmx,\bmy)$, but this is intractable owing to the curse of dimensionality and an infinite number of functions to choose from.
Instead, in {supervised learning}, one has labeled {training data} $\{\bmx_i, \bmy_i\}_{i=1}^N$ sampled from $p(\bmx ,\bmy)$.%
%% Based on this training data, supervised learning algorithms try to find the optimal function $\hat{f}$.
\footnote{Other machine learning paradigms like unsupervised, semi-supervised, and weakly supervised learning relax or remove the need for labels in the training data.}
Furthermore, the function space is restricted to a {model} -- a highly flexible family of functions $f_\phi(\bmx )$ parametrized by $\phi$. In this case, the algorithms minimize directly with respect to the model parameters $\phi$. Neural networks, support vector machines, and decision trees are examples of types of models commonly used in machine learning.  These models often have a large number of parameters, and in the case of neural networks, finding the optimal $f_\phi$ can be a difficult problem.
%% Modern machine learning has succeeded due to large amounts of training data, efficient training algorithms (see below) and fast hardware like GPUs.

An essential goal in machine learning is \emph{generalization}---the ability of the model to perform well on data which was not used in training. Failure in this task is called overtraining. There are a vast array of techniques to avoid overtraining (or overfitting) that can all be considered forms of {regularization}. Regularization techniques like dropout were key to advances in image recognition with deep learning~\cite{krizhevsky2012imagenet, imagenet-challenge}.  Theoretical analysis of the generalization of deep learning is difficult because it involves a complicated interaction between the specifics of the model $f_\phi$, the  optimization algorithms, the loss function, regularization techniques, and the specifics of the finite training samples and the true underlying distribution $p(\bmx ,\bmy)$.
Empirically, deep learning models generalize much better than existing theoretical analysis might suggest. While a more powerful theoretical analysis of generalization for deep learning would be valuable, in practical terms it is not necessary as long as statistically independent data, not used in training, is available to validate the performance.

\subsection{Neural Network Basics}

In the language of neural networks, the space of functions searched is defined by the structure of the networks, which defines a series of transformations. These transformations map the input $\bmx$ onto internal or ``hidden'' states $\bm{h}_i$, until the final transformation maps these hidden states onto the function output $\bmy$.

Mathematically, these transformation are expressed as

\begin{equation}
 \bm{h}_{i + 1} = g_{i}(W_{i} \bm{h}_{i} + \bm{b}_i)
\label{eq:dense}
\end{equation}

\noindent
where $g_i$ is a some function, called the activation function, and a particular $\bm{h}_i$ is the $i$-th transformation of the information in $\bm{x}$, called the embedding. In a simple case, the first embedding is simply the input vector $\bm{h}_0 \equiv \bmx$, and the final embedding is the output of the network. The elements of the matrix $W$ are referred to as {weights} and those of vector $\bm{b}$ as {biases}. The general structure of these transformations, such as the dimensionality of each $W$ and the choice of activation function is referred to as the network {architecture}, which, taken together with the training parameters constitute the {hyperparameters} of the network. 

The weights and biases of the network are initalized randomly. Finding the function which optimizes the loss function is done through an iterative process called training. Conceptually, this uses the labeled training examples $(\bmx, \bmy)$ and calculates the gradient of the loss function with respect to the model parameters, $\nabla_{\phi}L(f_{\phi}(\bmx ), \bmy)$. In practice, the calculations are done through a technique called backpropagation,
which is an efficient means of computing this gradient.
%% in which the difference between the function value $f_{\phi}(\bmx )$ and the desired output $\bmy$ is propagated back through the embeddings to understand how to adjust the parameters.
In principal, backpropagation puts only one restriction on $L$ and $f(\bmx)$: they must be be differentiable for a gradient to be defined. %% In practice, backpropagation is difficult to implement efficiently and is rife with suitabilities and numerical pitfalls. A pragmatic physicist would be well advised to leave implementation details to one of many existing open source libraries~\cite{chollet2015keras,tensorflow,theano}.

\subsection{Deep Learning}

Initially, the term {deep neural networks} referred to neural
networks with many hidden layers, and it was used to differentiate
such networks from {shallow neural networks}, which had only
one hidden layer. For many years, it was argued that using a
shallow network was not a restriction, because of the theoretical
analysis that demonstrated that any function can be approximated by a shallow
network~\cite{hornik_multilayer_1989}. However, an effective
shallow network may require an enormous number of nodes in the
hidden layer, and in practice, shallow neural networks often
failed to discover useful functions from high-dimensional
data sets.

The traditional strategy for discovering the optimal function for a given application involves a gradient search through $f_\phi$. In practice, this becomes much more difficult to accomplish as the neural network becomes deeper. As the difference between the function value $f_{\phi}(\bmx )$ and the desired output $\bmy$ is propagated back through the various embeddings, the gradient $\nabla_{\phi}L(f_{\phi}(\bmx ), \bmy)$ rapidly approaches zero, making it difficult to improve the performance by adjusting the model parameters. This {vanishing gradient problem} \cite{hochreiter_recurrent_1998,bengio1994learning} has been overcome in recent years using a variety of strategies, including computational boosts from graphical processing units, larger training samples, new regularization techniques such as dropout \cite{dropout2012}, pre-training of initial embeddings with unsupervised learning methods such as autoencoders \cite{hinton_fast_2006, bengio_greedy_2007}. Autoencoders attempt to learn a useful layered representation of the data without having to backpropagate through a deep network; standard gradient descent is only used at the end to fine-tune the network.   

More generally, {deep learning} can refer to a broad class of
machine learning methods emphasizing hierarchical representations
of the data and modular, differentiable components. Not only do
these deep networks have more expressive capacity, but also the layers
can be interpreted as building up a hierarchical representation of
the data. In natural images, for example, the first layers learn
low-level features like edges and corners, the middle layers learn
midlevel features like eyes, and the final layers learn high-level
features like faces. The processes that produce particle physics
data naturally lead to compositionality and hierarchical
structured data. For instance, a typical event at the LHC is
composed of jets, jets are composed of hadrons, hadrons lead to
tracks and calorimeter clusters, tracks are composed of hits, and
calorimeter clusters are composed of calorimeter cells. The
analogy also extends to higher levels with groups of particles
forming resonances in a cascade decay. For these reasons, one
might anticipate deep learning to be particularly effective at the
LHC.

Modern deep learning is characterized by the composition of
modular, differentiable components~\cite{Goodfellow-et-al-2016}.
Among the first of these modular components was the convolutional
filter, which is arguably the most important innovation in deep
learning applied to image processing~\cite{convnet}. Convolutional
architectures are natural when the input data has some notion of
locality, the individual components of $\bm{x}$ are the same type
(e.g., neighboring pixels in an image), and the interesting
features are equally likely to appear in any local patch. The
kernel $\bm{k}$ of the convolution can be interpreted as a bank of
filters that operates on a local patch of the input $\bm{x}_i$ as

\begin{equation}
\bm{h} = g(W \bm{x} + \bm{b}) \quad \to \quad h_{i,j} = g(\bm{k}_j
\cdot \bm{x}_{i} + b_j) \; , \label{eq:conv}
\end{equation}
where $i$ indexes the local patches and $j$ indexes the filters.
Because the same kernel is applied as it is swept over the input,
it has the effect of sharing weights in a dense network. Weight
sharing imposes translational symmetry on the network, and it
drastically reduces the number of parameters in the network and
the amount of data needed to train them. Convolutional layers are
usually followed by a pooling layer, which summarizes the result
of applying the filters in a local patch (e.g., by taking the
maximum or average).  These convolutional and pooling layers can
be composed to build a hierarchical representation of the data
going from low- to mid- to high-level features. Other modular components
include normalization layers~\cite{ioffe2015batch} and residual
layers~\cite{he2016deep}.  By training the different layers of
these networks jointly, deep convolutional neural networks learn
hierarchical features that tend to outperform engineered features
for image processing tasks.

\begin{figure}
\includegraphics[width=\textwidth]{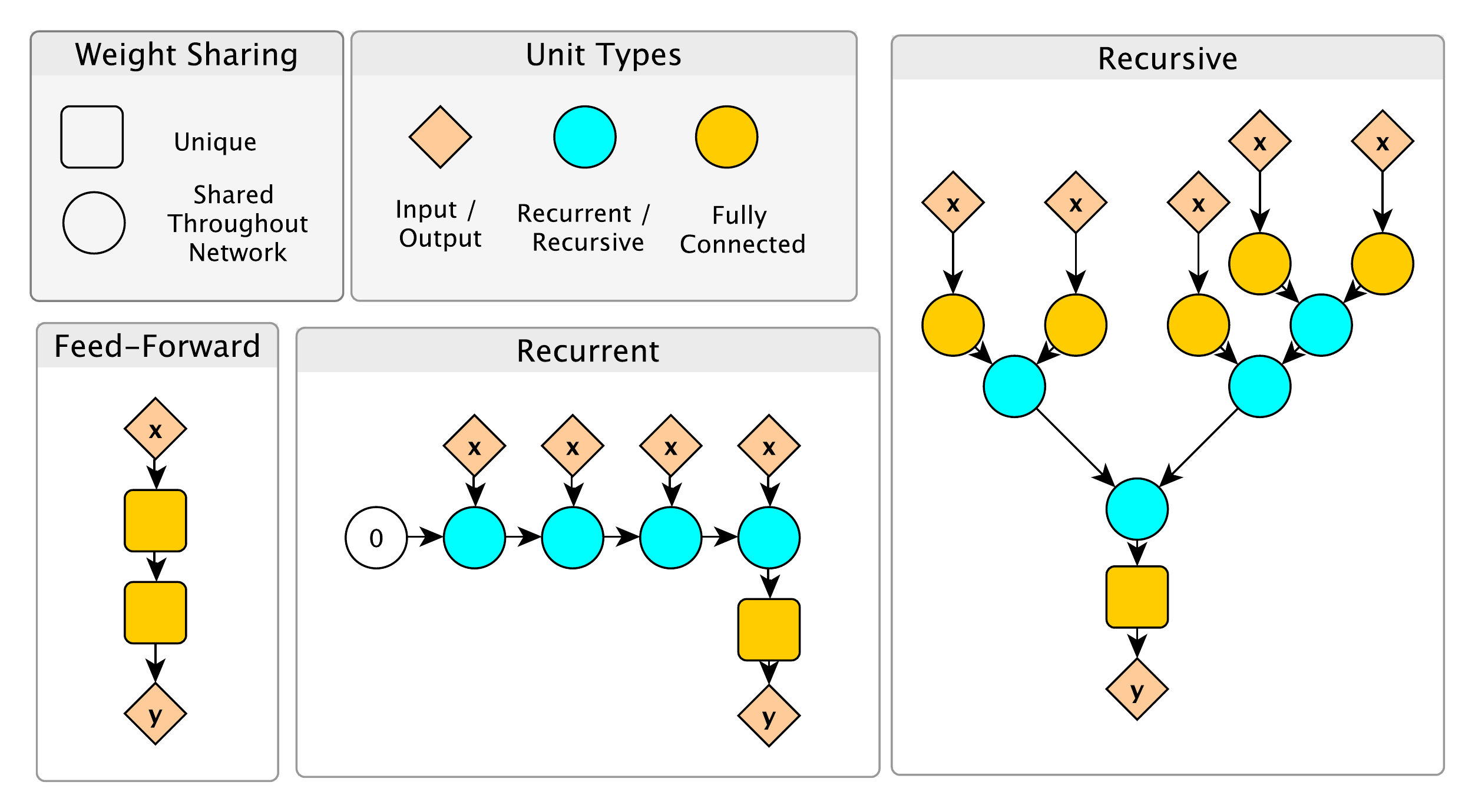}
\caption{Schematic showing feed-forward, recurrent, and recursive neural network architectures. Diamonds represent inputs and outputs, while processing units are represented with circles and squares. Arrows between processing units represent embeddings $\bm{h}$. Standard feed-forward networks map a fixed length $\bmx$ into $\bmy$, whereas recurrent and recursive networks can process a sequence of inputs $\{\bmx_i\}$.
Units represented as circles are shared throughout the network: once the network is trained, the units can be used to build a network of arbitrary size.
Recurrent networks can be viewed as a subset of recursive networks, in which each node combines one input $\bmx_i$ and the output from the previous recurrent node $\bm{h}_{i-1}$ to produce $\bm{h}_i$, and where $\bm{h}_0 = 0$.
Recursive units map each pair of inputs to an output in the same space, $(\bm{h}_i, \bm{h}_j) \to \bm{h}_k$.
%% While in principal the raw inputs $\bmx$ could serve as the leafs in the recurrent tree, it's often useful to map each $\bmx \to \bm{h}$ as a first step.
Note that these components can also be chained: Any
output node can also serve as an input node to another component.
} \label{fig:networks}
\end{figure}

Working with variable-length input (e.g., words in a sentence)
requires the network architecture to be adaptive in some way.
Variable-length input can be cropped or padded with zeros to fit into a fixed-size vector $\bmx$, but
these blunt solutions either discard potentially useful
information or force a network to accommodate placeholder values.
A far more natural solution is to rely on networks that can
adjust to the input size dynamically. A particularly
illustrative case is a simple recursive unit, which maps a pair of
inputs, $\bm{h}_1$ and $\bm{h}_2$, onto an output,
$\bm{h}$, as follows:

\begin{equation}
\bm{h} = g_{i}(W_1 \bm{h}_1 + W_2 \bm{h}_2 +
\bm{b}). \label{eq:recurrent}
\end{equation}
Assuming that one or both of the input vectors are of the same
dimension as $\bm{h}$, the output can be fed into the input
recursively and condense an arbitrary length sequence of inputs
into a fixed-dimensional representation $\{\bm{h}_i\} \to
\bm{h}$. More generally, a neural network can be visualized as
a directed acyclic graph in which edges represent the various
internal $\bm{h}$ vectors. Figure~\ref{fig:networks} illustrates several such graphs.

In practice, since recursive networks can grow very deep, simple recursive units encounter problems with vanishing or exploding gradients. These longer sequences can be handled using a technique known as gating, where activation functions and transformations are applied selectively, or inputs can be ignored entirely. These alleviates the exploding and vanishing gradient problem at the expense of a more complicated recurrent unit; examples are long-short-term-memory (LSTM) units~\cite{hochreiter1997long}, and gated recurrent units (GRU)~\cite{cho2014learning}. 

Both convolutional and recurrent layers are examples of network
architectures that use {shared weights}. In the
convolutional case, each element of $\bm{k}$ acts in multiple dot
products, whereas in the recurrent case, the transformation in
Equation \ref{eq:recurrent} is applied multiple times for each pattern.
Weight sharing can be viewed as a type of regularization: by
reusing the same transformation in multiple places throughout the network, the
network designer can encode domain-specific structure.

%% The growing complexity of neural networks, in terms of both the
%% architecture and the raw computational power required for
%% training, might seem overwhelming to the pragmatic high-energy
%% physicist. Worse, the summary above is merely a brief review of
%% state-of-the-art deep learning, and given the rapid pace at which
%% the field evolves, techniques like gating may change in the near
%% future.  Fortunately, the basic principles of deep
%% learning---computing the gradient and updating the internal
%% parameters to minimize it---are unchanged by the addition of more
%% gates or more complicated graphs. Several software packages,
%% providing automatic differentiation, efficient caching, and
%% interfaces to hardware such as GPUs,  significantly simplify the training
%% procedure and are well supported outside HEP. Thanks to these tools, the role of the physicist is reduced
%% to choosing an appropriate architecture and training strategy.

\section{SURVEY OF APPLICATIONS}

Machine learning has found numerous natural applications
in particle physics, where many tasks require classification in
high-dimensional variable spaces. At the lowest level, machine learning tools
can perform hit reconstruction~\cite{Aad:2014yva} or
track finding~\cite{Peterson:1988gs} in individual detector
systems. These tools can also perform object identification by using
information from various detector systems, such as
electron~\cite{Abramowicz:1995zi},
photon~\cite{cms2014observation}, or
$\tau$ lepton~\cite{Abazov:2004vd} identification. Finally, machine learning
tools have been widely used to classify entire events as
background-like or signal-like, both in the final statistical
analysis~\cite{ABREU1992383} or at the initial trigger
decision~\cite{Kohne:1997ph}.  These machine learning tools have found
high-profile application in single \textit{t} quark
searches~\cite{Abazov2001282}, early Higgs boson
searches~\cite{Aaltonen:2009jg}, and the Higgs boson
discovery~\cite{cms2014observation}.

\subsection{Event Selection and High-Level Physics Tasks}\label{sec:apps}

The earliest successes of deep learning in high energy physics came in improvements in event selection for signal events with complex topologies. In the past few years, several studies have demonstrated that the traditional shallow networks based on physics-inspired engineered (``high-level'') features are outperformed by deep networks based on the higher-dimensional features which receive less pre-processing (``lower-level'') features. Prior to the advent of deep learning, such pre-processing was necessary, as shallow network performance on low-level features fell short. The deep learning results discussed below demonstrate that deep networks using the low-level features surpass the shallow networks using high-level features. This confirms the suspicion that feature engineering, applying physics knowledge to construct high-level features, is often sub-optimal.

\begin{figure}
  \begin{minipage}{.5\textwidth}
    \includegraphics[width=\textwidth]{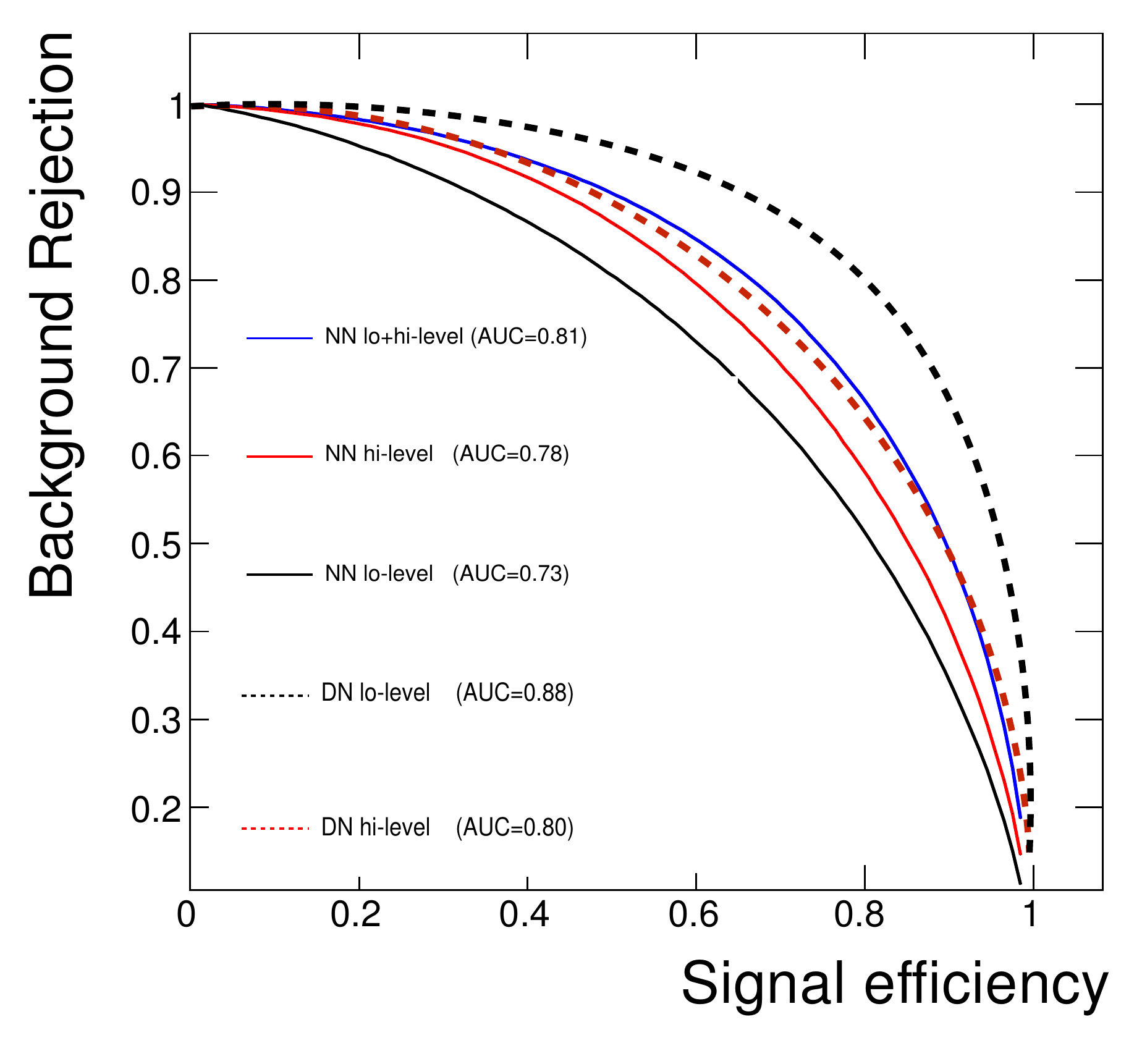}
  \end{minipage}
  \begin{minipage}{.5\textwidth}
\includegraphics[width=\textwidth]{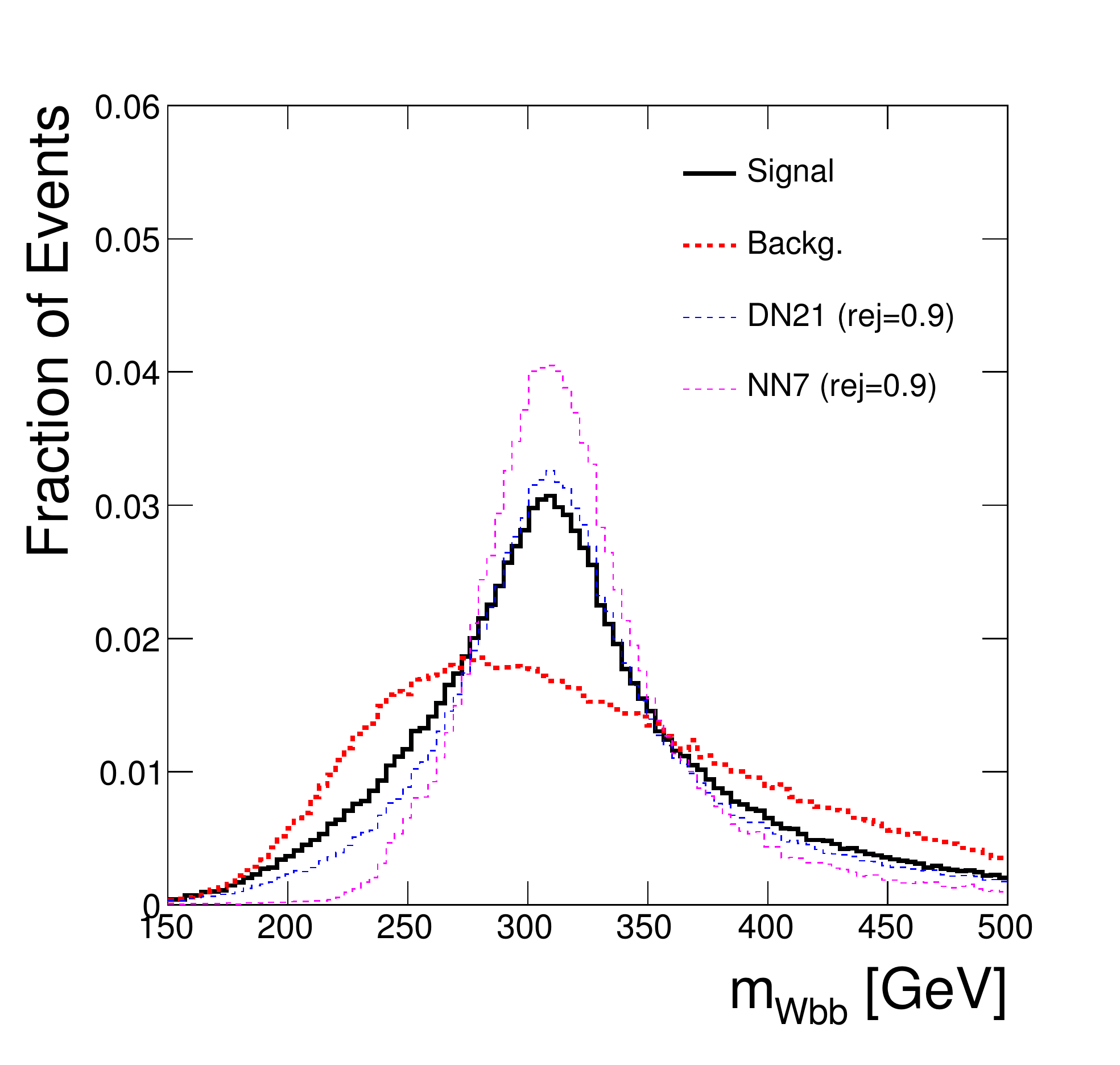}
  \end{minipage}
\caption{ (left) Deep networks (DN) performance in signal-background classification compared to shallow networks (NN) with a variety of low- and high-level features demonstrate that deep networks with only low-level features outperform all other approaches, from \cite{Baldi:2014kfa}. (right) Comparison of the
distributions of invariant mass of events selected by a deep
network (DN21) using only object momentum to a shallow network
(NN7) that has been trained using this feature, at equivalent
background rejection. Also shown are the distributions in pure
signal and background samples. The shallow network relies heavily
on this invariant mass quantity to discriminate. The deep network
has also discovered the value of this feature, but is able to
recover signal further from peak.}
\label{fig:dnn_mass}
\end{figure}

An early study~\cite{Baldi:2014kfa} compared the performance of
shallow and deep networks in distinguishing a cascading decay of
new exotic Higgs bosons from the dominant background. This study used a structured
data set in which a large set of basic low-level
features (object four-momenta) were reduced to a smaller set of
physics-inspired high-level engineered features.   Because the
high-level features were a strict function of the low-level
features, they contained a subset of the information, so that the
expertise encoded by the high-level features was solely in the
design of these dimensionality-reducing functions rather than the
introduction of new information.  This gave rise to revealing
comparisons about the relative information content of the low- and
high-level features and the power of classifiers to extract it. In
their study, Baldi {et al.}~\cite{Baldi:2014kfa} found that deep
networks using the lower-level data significantly outperformed
shallow networks that relied on physics-inspired features such as
reconstructed invariant masses (Figure~\ref{fig:dnn_mass}). The
high-level engineered features captured real insights, but clearly
sacrificed some useful information.

Such conclusions are not universal, however, but rather are dependent on the
specifics of the classification task. Using the same approach,
Baldi {et al.}~\cite{Baldi:2014kfa} analyzed a supersymmetric
particle search that has received significant feature engineering
in the literature, and found that shallow networks using low-level
data very nearly matched the performance of both shallow networks
using engineered features as well as deep networks on either set
of features. The authors concluded that this application requires only
simple linear functions on the lower-level data, and may not
require a deep network or deserve such attention to feature
engineering.  Similar conclusions were reached in
Reference~\citen{Santos:2016kno}.

In addition to optimizing event selection for a fixed signal
versus background problem, Cranmer {and colleagues~\cite{Louppe:2016aov}} showed that it is
possible to approximate the likelihood ratio $p(\bmx
|\theta)/p_0(\bmx )$ for a continuous family of signal models
parameterized by $\theta$. Since binary classification
amounts to approximating a likelihood ratio, this generalization
is called a {parameterized classifier}. This is a common-use
case at the LHC because most signals are predicted by theories with
several free parameters. For instance, Baldi {et al.}~\cite{Baldi:2016fzo} used this technique to create a classifier for $X\to t\bar{t}$ versus
Standard Model background parameterized by the mass of the
resonance $m_X$.  The approximate likelihood
can also be used in a likelihood fit to estimate the parameters
$\theta$ (e.g., masses, coupling constants), providing a
novel form of likelihood-free inference. The \textsc{carl}
software package provides a convenient interface for this
technique~\cite{Louppe:2016aov}.

\subsection{Jet Classification}

Machine learning has been applied to a wide range of jet classification problems, to identify jets from heavy ($c$, $b$, $t$) or light ($u$, $d$, $s$) quarks, gluons, and $W$, $Z$, and $H$ bosons. Traditionally these classification problems have been grouped into flavor tagging, which discriminates between $b$, $c$, and light quarks, jet substructure tagging, which discriminates between jets from $W$, $Z$, $t$ and $H$, and {quark--gluon tagging}. 

In flavor tagging, the discriminating information
is spatial: Heavy quarks decay weakly in a matter of picoseconds,
which is sufficient time for a highly boosted quark to travel
roughly a centimeter from the interaction point. Due to this
measurable separation, flavor tagging relies heavily on tracks
reconstructed by high-granularity sensors near the interaction
point, and on vertices fit to these tracks. The use of machine
learning in flavor tagging dates to LEP~\cite{dg95}, where
libraries such as JETNET~\cite{jetnet} were used to identify
\textit{b} and \textit{c} quarks, and has continued through LHC runs 1 and
2~\cite{CMS:2016kkf,Aad:2015ydr,ATL-PHYS-PUB-2017-013}.

In contrast to jet flavor tagging, jet substructure and
quark--gluon tagging rely on information created at one spatial
location during the decay of the original particle. The spread in
decay product momenta translates to spatial separation as the
particles travel away from the interaction point, but the
underlying physics is localized at the decay point. Thus, while the
power of flavor tagging is limited primarily by the tracking
detector resolution, jet substructure and quark--gluon
discrimination are subject to quantum-mechanical limitations.
Theoretical and experimental physicists have expended considerable
effort in quantifying these limitations, and in engineering
jet-substructure-based discriminating
variables~\cite{Larkoski:2017jix}.

\begin{figure}
  \begin{minipage}{.5\textwidth}
    \includegraphics[width=\textwidth]{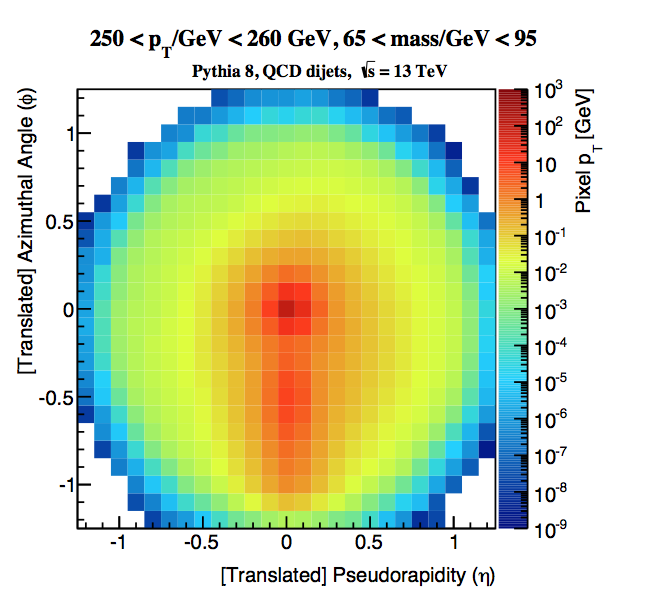}
  \end{minipage}
  \begin{minipage}{.5\textwidth}
    \includegraphics[width=\textwidth]{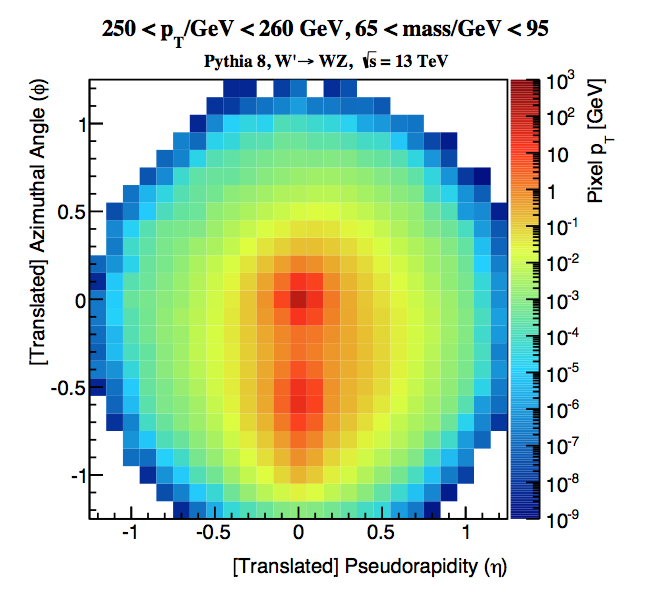}
  \end{minipage}
\caption{Example jet image inputs from the jet substructure classification problem described in Ref.~\cite{deOliveira:2015xxd}.  The background jets (left) are characterized by a large central core of deposited energy from a single hard hadronic parton, while the signal jets (right) tend to have a subtle secondary deposition due to the two-prong hadronic decay of a high-$p_{\textrm{T}}$ vector boson. Use of image-analysis techniques such as convolutional neural networks allow for powerful analysis of this high-dimensional input data.}
\label{fig:images}
\end{figure}

Recently, the realization that lower-level, higher-dimensional
data could contain additional power led to a rapid proliferation
of studies that challenged established substructure
approaches~(\citen{Larkoski:2017jix}; see https://indico.physics.lbl.gov/indico/event/546/overview). In 2014, Cogan {et
al.}~\cite{Cogan:2014oua} recognized that the projective tower
structure of calorimeters present in nearly all modern HEP
detectors was similar to the pixels of an image (Figure~\ref{fig:images}). This representation of the data allowed
physicists to leverage the advances in image classification such
as convolutional neural networks. The image-based networks
discriminated as well as or better than shallow networks using jet-substructure-based
inputs~\cite{Almeida:2015jua,deOliveira:2015xxd}. The
discriminating persisted in the presence of pileup and jet
grooming~\cite{Baldi:2016fql}, across
generators~\cite{Barnard:2016qma}, and could be generalized to
three-dimensional detectors using multiple stacked channels
analogous to colors~\cite{Komiske:2016rsd}. Outside collider
physics, a similar approach was used to tackle
object-identification tasks in neutrino
experiments~\cite{Acciarri:2016ryt,Aurisano:2016jvx}.

While the image-based approach has been successful, the actual
detector geometry is not perfectly regular; thus, some
preprocessing is required to represent the jet as an image. In
addition, jet images are typically very sparse. The sparsity can be
alleviated by enlarging pixels, but the harsher discretization
sacrifices resolution in $\eta$ and $\phi$. Given that the jets
themselves are composed of a varying number of reconstructed
constituents, each with well-defined coordinates and parameters,
jet tagging algorithms that can work with a
variable number of inputs are desirable.

\begin{figure}
  \begin{minipage}{.6\textwidth}
\includegraphics[width=\textwidth]{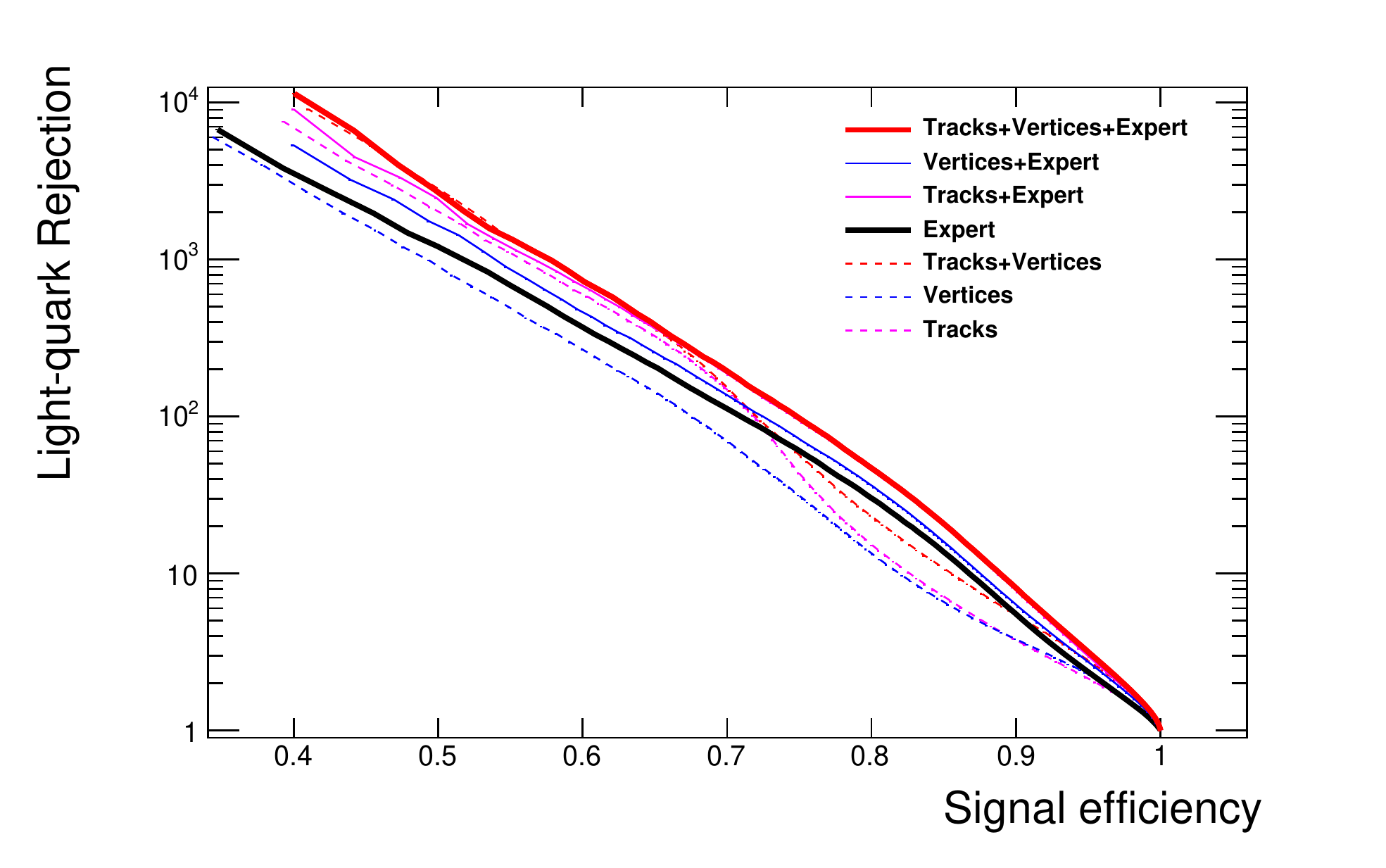}
  \end{minipage}
  \begin{minipage}{.4\textwidth}
\includegraphics[trim={0  3cm 0 3cm},clip,width=\textwidth]{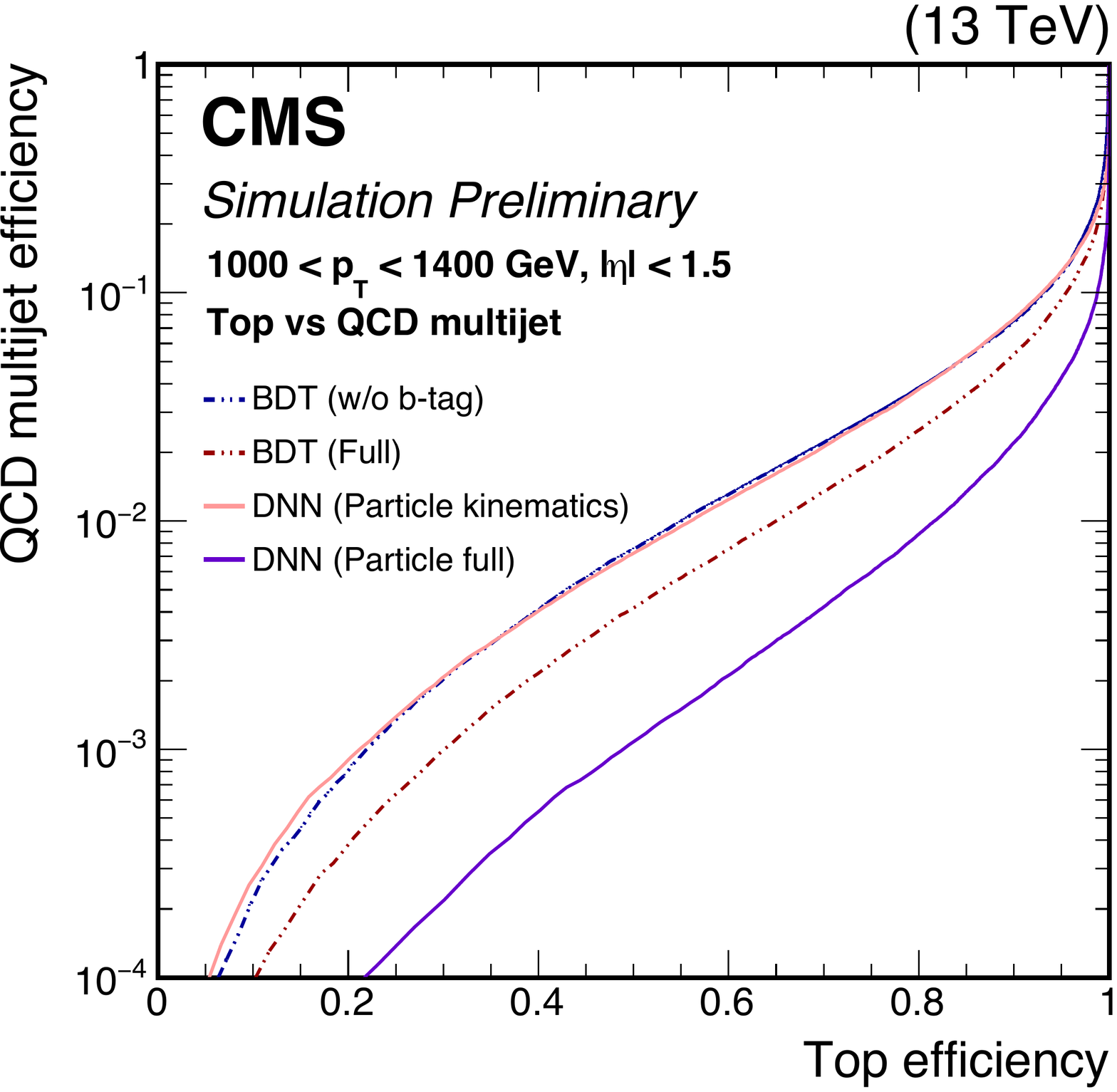}
  \end{minipage}
\caption{(\textit{left}) A comparison~\cite{Guest:2016iqz} of the
jet-flavor-tagging performance of deep networks with varying
levels of feature engineering.  The lowest-level data (labeled Tracks)
contain all of the information and perform well, but networks
that are also given vertices or higher-level features
(labeled Verticies and Expert) still show some improvement. (\textit{right}) A
comparison~\cite{CMS-DP-2017-049} of the performance of two boosted decision tree (BDT)
taggers and two particle-based deep neural network (DNN)\ taggers in
simulated events of four top jets as signal and QCD multijets as
background.} \label{fig:flavor}
\end{figure}

Several flavor-tagging applications have made use of deep networks
trained on variable length arrays of track parameters. Guest {et
al.}~\cite{Guest:2016iqz} investigated the need for
feature-engineering by defining {low-level},
{mid-level}, and {high-level} features, where the
mid- and high-level features were inspired by typical
flavor-tagging variables and derived from a strict subset of the
low-level feature information. The authors found similar discrimination
using fixed-size, zero-padded networks and recurrent
architectures, and that the best performance came from using all
three levels of features (Figure~\ref{fig:flavor}).
Both ATLAS and CMS have since commissioned flavor-tagging neural
networks that rely on individual tracks or, in the CMS case,
particle-flow candidates. The ATLAS recurrent-network-based
approach reduces backgrounds by roughly a factor of two when
combined with traditional high-level
variables~\cite{ATL-PHYS-PUB-2017-003,ATL-PHYS-PUB-2017-013}.
CMS's \textsc{DeepFlavor}~\cite{CMS-DP-2017-012,CMS-DP-2017-005}
neural network first embeds each flow candidate with a
transformation that is shared across candidates, then combines
the candidates' high-level variables in a single zero-padded dense
network.

Networks trained on variable-length arrays of jet constituents
proved equally useful in boosted top tagging. In one series of
studies, a zero-padded dense network showed
promise~\cite{Pearkes:2017hku}, but backgrounds were halved by
replacing the dense network with a recurrent
network~\cite{Egan:2017ojy}. The CMS Collaboration experimented
with two variants of the \textsc{DeepJet}~\cite{DeepJetNIPS}
algorithm. The first was similar to \textsc{DeepFlavor}, whereas
the second replaced the dense network with a recurrent neural network. In comparison to a
baseline that combined high-level variables in a boosted decision
tree, QCD\ multijet misidentification was reduced by a factor of approximately four at 60\% top-tagging
efficiency (Figure~\ref{fig:flavor})~\cite{CMS-DP-2017-049}.

\begin{figure}
  \includegraphics[width=0.9\textwidth]{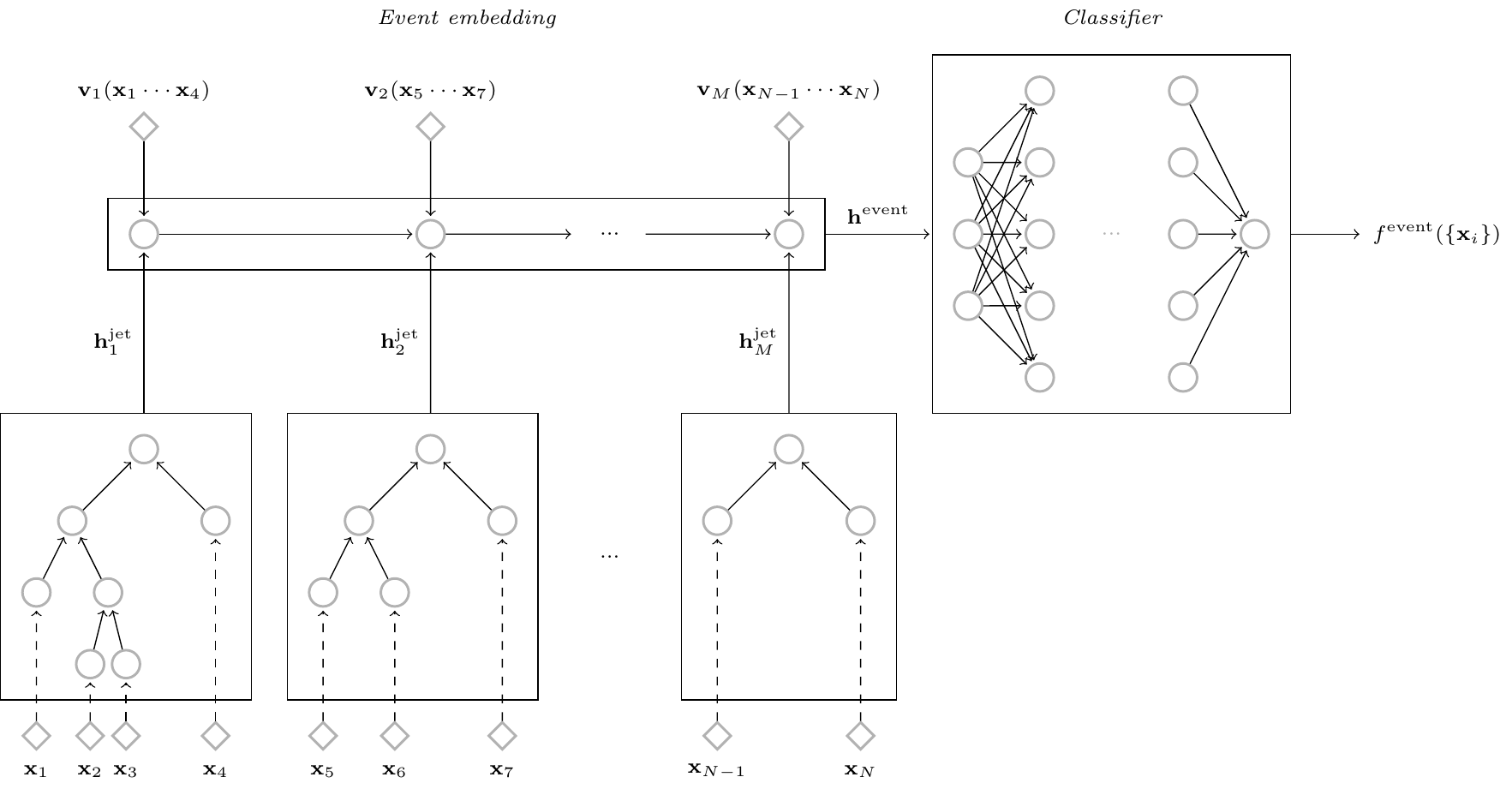}
\caption{A schematic showing the
hierarchical composition of a deep learning model following the
outline of traditional high-energy physics data pipeline. The lowest-level
detector inputs are represented by $\bmx_i$ (\emph{diamonds}), which are then fed into recurrent
networks (\textit{lower boxes}) to form jet embeddings
$\bm{h}_1^\textrm{jet}$. These are augmented by jet-level features $\bm{v}_i$. The jet
embeddings are processed by a network to form a final event-level
embedding  $\bm{h}^\textrm{event}$, which is
then fed into a classifier, leading to the output
$f^\textrm{event}(\{\bmx_i\})$. The entire network can
be learned jointly, or individual components can be pre-trained. Adapted from~\cite{Louppe:2017ipp}.}
\label{fig:embedding}
\end{figure}

Recurrent networks act on sequences, requiring an ordering of
the particles. While several natural orderings exist, the most
natural is arguably the $k_{\rm T}$ jet clustering history, which
defines a tree that can be used as the scaffolding for a recursive
neural network. Louppe {et al.}~\cite{Louppe:2017ipp} applied a recursive neural network
over the jet clustering history, providing a hybrid QCD-aware
neural network strategy. In the same $W$-versus-QCD jet classification problem as studied in
Reference~\citen{Barnard:2016qma}, the recursive network showed no
improvement over image-based networks when trained on jet images,
but improved substantially when the image preprocessing was
removed. Notably, the recursive and recurrent networks had fewer
parameters and thus required far fewer data to train. The same
recursive neural network has been applied to quark--gluon
tagging~\cite{Cheng:2017rdo}.

The clustering of objects
need not end at the jet level (Figure~\ref{fig:embedding}). The outputs from the jet-level
recurrent network can be fed into a recurrent network to produce a
high-level event embedding.

More recently, Henrion \textit{et. al} investigated representing jets as a graph instead of as a tree~\cite{jetsAsGraphsNIPS}. Graph convolutional networks provide a generalization of convolutional neural networks that can be applied to irregularly sampled data~\cite{DBLP:journals/corr/BronsteinBLSV16}. In this picture, the particles represent the nodes of the graph and the edges can encode how close the particles are in a learned adjacency matrix. Henrion  \textit{et. al} showed that such a network outperformed a recursive network for the same $W$ vs. QCD  jet tagging problem studied in Refs.~\cite{Barnard:2016qma, Louppe:2017ipp}.

\subsection{Tracking}

Track-reconstruction algorithms are among the most central processing unit (CPU) and data
intensive of all low-level reconstruction tasks. The initial stage
of track reconstruction involves finding {hits}, or points
where some charge is deposited on a sensing element. In the case
of the pixel sensors that form the innermost layer of the
detector, neighboring hits are clustered into {pixel
clusters}, which then form track {seeds}. These seeds form a
starting point for a Kalman filter, which extends the seeds into
full {tracks} that extend to the calorimeters. The entire
procedure can be viewed as a sequence of clustering algorithms,
in which the zero-suppressed readout from $\mathcal{O}(10^8)$ channels
provides $\mathcal{O}(10^4)$ hits, which are then clustered into $\mathcal{O}(10^3)$
tracks per event.

Machine learning has proven useful in several aspects of track
reconstruction. In cases where multiple tracks pass through the
same pixel cluster, ATLAS relies on neural networks to return a
measurement for each track rather than assigning each to the
cluster center~\cite{Aaboud:2017all,Salzburger:2018442}. LHCb
makes use of several shallow networks in track reconstruction as
well. Due to the long distances between sensor elements in the
LHCb tracker, falsely connected hits forming ``ghost tracks'' are
a large source of backgrounds. A simple three-layer network
reduces this background by a factor of two in comparison to a
$\chi^2$-based discriminant~\cite{Stahl:2260684}. Several other
networks are evaluated to filter out fake tracks before running a
full track fit or to recover tracks with missing hits.

Thanks to these algorithms and careful tuning, track reconstruction
is nearly 100\% efficient and spuriously reconstructed tracks are
rare, meaning that the clustering aspect of tracking is largely solved.
Reducing the CPU overhead remains a significant problem, however,
especially within high-level trigger farms. Within ATLAS and CMS,
these are clusters of $\mathcal{O}(10^4)$ processors that must reconstruct
$\mathcal{O}(10^5)$ events per second~\cite{MachadoMiguens:2230025}. To keep
tracking CPU costs manageable, the experiments reconstruct
tracks only in limited regions of the detector. These regions are
selected on the basis of their proximity to muons or to calorimeter energy deposits that are
consistent with relatively rare physical signatures like leptons
or high-$\pt$ jets. While effective, this selective tracking
severely hampers searches and physical measurements that rely on
low-$\pt$ track-based signatures.

These issues will be compounded considerably in the high-luminosity LHC (HL-LHC). The majority of the tracking CPU budget is
currently allocated to track-building phase, which depends on an
expensive Kalman filter~\cite{Adam:2003kg,ATLAS-CONF-2012-047}.
With the higher track densities expected at the HL-LHC, the number
of false seeds is expected to increase
combinatorially~\cite{tracking.hllhc}, as is the probability of
the Kalman filter building a branching track. More-discriminating
seeding algorithms or trainable deep Kalman
filters~\cite{deepkalman1,deepkalman2} could help reduce the
number of track-fitting iterations.

Unfortunately, most tracking software is deeply interwoven with
the experiments' reconstruction frameworks and, as a result, is
poorly suited for the quick exploratory studies that will be
needed to develop the next generation of algorithms.

In
anticipation of the coming HL-LHC data onslaught, however,
improved tracking is essential. Several projects aim to accomplish
this goal indirectly by increasing the visibility and
accessibility of tracking software. Most ambitiously, the ACTS
project~(see http://acts.web.cern.ch/ACTS/index.php, https://github.com/trackml) seeks to implement a modular and
experiment-independent software stack for tracking-related
studies, including detector geometry, an event data model, and
seed-finding and track-fitting tools. Simplified tracking
models~\cite{heptrkx} have been used as a basis for studies
showing the viability of LSTMs for track
building~\cite{trackingNIPS} or for tracking data
challenges~\cite{track.ramp}.

\subsection{Fast Simulation}

The ability to model high-dimensional distributions not only enables improved statistical analysis, but also provides a new path towards fast simulation. Fast simulation is valuable because the full
simulators, which faithfully describe the low-level interactions
of particles with matter, are very computationally intensive and
consume a significant fraction of the computing budgets of current
experimental collaborations.

Until now the dominant approach to fast simulation has been to develop fast parametric distributions largely by hand~\cite{ATLAS:2010bfa}.  A more recent deep-learning approach is to train a network to learn the simulation from an initial pool of traditionally simulated events.  This approach creates a {generative model} $G$ which which approximates the distribution of samples produced by the simulator by mapping an input space of random numbers to the space of the data.

One promising approach is based on Generative Adversarial Networks (GANs). The training of the generative model $G$ is accomplished through competition with an adversary network $A$. While $G$ generates simulated samples, the adversarial network $A$ tries to determine whether a given sample is from the generative model or from the full simulator.  The two networks are pitted against each other: $A$ attempts to identify differences between the traditional samples and those generated by $G$, while $G$ attempts to fool $A$ into accepting its events, and in doing so learns to mimic the original sample generation. The stability of such a training arrangement, however, can be difficult to achieve, and expert knowledge is required to construct an effective network.

Paganini {\it et al.}~\cite{Paganini:2017hrr} applied the GAN approach to simulation of the electromagnetic showers in a multi-layer calorimeter, one of the most computationally expensive steps for the low-level simulator. They report large computational speedups while achieving reasonable  modeling of the energy deposition, though not yet matching the accuracy of the full simulation. In a related approach, de Olivereira {\it et al.} find similar success simulating jet images~\cite{deOliveira:2017pjk}.  Future simulation tools built on GANs may provide important speed boosts for the slower elements of the simulation chain, or they may be sophisticated enough to provide end-to-end simulations.

The resulting network evaluation is much less computationally demanding than the low-level simulation, and can be viewed as a non-parametric fast simulation. The promise of this approach to mitigate the computational burden for simulation has been called out in the strategic planning for HL-LHC software efforts~\cite{Elmer:2017rej, Alves:2017she}. See Refs. ~\cite{larochelle2011neural,2015arXiv150505770J,DBLP:journals/corr/DinhKB14,Cranmer:2015bka,kingma2016improving, papamakarios2017masked} for alternative approaches to fast simulation.

\subsection{Impact}

Taken together, the new tools made possible by deep learning promise to make a significant impact on high-energy physics. The specific examples above -- mass resconstruction, jet substructure and jet-flavor classification -- are important benchmarks and long-standing challenges. The significant improvements offered by deep learning in these areas support the claim that many areas of LHC data analysis suffer from long-standing sub-optimal feature engineering, and deserve re-examination.

\section{CONCERNS}

\subsection{What Is the Optimization Objective?}

A challenge of incorporating machine learning techniques
into HEP data analysis is that tools are often optimized for performance on a particular task that is several steps removed
from the ultimate physical goal of searching for a new particle or
testing a new physical theory. Moreover, some tools are used in
multiple applications, which may have conflicting demands. For
instance, a deep learning jet flavor-tagging algorithm might be
used for searches for supersymmetry as well as precision
measurements of the Higgs sector, which may have different needs
with respect to balancing signal efficiency and background
rejection.

These considerations are further complicated by the fact that\
the sensitivity to high-level physics questions must account for
systematic uncertainties, which involve a nonlinear trade-off
between the typical machine learning performance metrics and the
systematic uncertainty estimates. For example, a new classifier
may have a better false-positive rate than a baseline algorithm,
yet simultaneously be more susceptible to systematic mismodeling
between the simulation and the real data. Whether or not this new
classifier will improve the sensitivity for the ultimate
high-level physics goal depends on details such as the
signal-to-background ratio, the total number of data, and the size
of the systematic uncertainty, which are not typically included in
the classifier training.

Traditionally, HEP physicists have taken these
considerations into account through heuristics and intuition. But as deep
learning penetrates  into the analysis pipeline, it is important to
revisit these trade-offs and attempt to make them explicit to
design new loss functions and learning algorithms that directly
optimize for our ultimate physics goals.

For example, in order to to be robust to systematic uncertainties, one can use a  classifier parametrized in terms of the nuisance parameters~\cite{Cranmer:2015bka,Baldi:2016fzo}, allowing for major speedups compared to earlier strategies~\cite{Aaltonen:2008bd}. An alternative approach is to train a network to be insensitive to
the systematic uncertainty, which is achieved either by
boosting~\cite{Stevens:2013dya} or by using an adversarial
training procedure that encourages the output of the network to be
independent of the nuisance
parameters~\cite{2014arXiv1412.4446A,Louppe:2016ylz}. The
technique can also be used to enforce independence from another
variable, such as the jet mass~\cite{Shimmin:2017mfk}. In general,
there is  a trade-off between performance in the classification (or
regression) task and robustness to systematics, which can be
adjusted via a  hyperparameter $\lambda$ in the objective
function. Unfortunately, optimization of $\lambda$ requires
retraining the network, and the objective function may not be
differentiable with respect to $\lambda$.

Optimization of differentiable components is efficiently handled
with various forms of stochastic gradient descent, although these
algorithms often come with their own hyperparameters. The
optimization with respect to hyperparameters that arise in the
network architecture, loss function, and learning algorithms are
often performed through a black-box optimization algorithm that
does not require gradients. This includes Bayesian
optimization~\cite{spearmint, sckopt} and genetic
algorithms~\cite{Aaltonen:2008bd}, as well as variational
optimization~\cite{vopt, avo}.

\subsection{Interpretability and Reliance on Simulation}

Machine learning provides an effective and powerful solution to many data analysis challenges in HEP, in some cases lessening the need for engineered features driven by physical insight. However, in some sense this approach is not satisfactory, as progress on the computational side is not always matched by gains in physical understanding and is heavily reliant on simulation programs.

The nonparametric nature of the neural network approach makes it
very difficult to interpret the solution. Unlike a simple analytic
function written in mathematical language familiar to physicists,
a neural network cannot be easily inspected to discover the
structure of its learned solution.  Due to the high-dimensional
nature of the input data $\bmx$, reverse-engineering the
classification strategy to identify signal-like or background-like
regions of the original space is also very
challenging~\cite{Chang:2017kvc}. This is not surprising, and
generically we should anticipate a trade-off between performance
and interpretability.

Aside from understanding the nature of the decision, the use of
machine learning also complicates the scientific communication and
theoretical interpretation of LHC results. While traditional
cut-based analyses can be conveyed in tables and prose, a learned
neural network cannot. Not only does this make it difficult for a
reader to glean the essential physics, it is also an
impediment to reinterpreting the result in the context of a
different theoretical model. This issue further motivates analysis
preservation and reinterpretation systems, such as
RECAST~\cite{Cranmer:2010hk,Cranmer:2017frf}, that can re-execute
the original data analysis pipeline.

Systematic uncertainties due to mismodeling in the simulation are
also a major concern. The networks are routinely trained using
large samples of simulated collisions, and the nature of their
solution is relevant to experimental data only if those simulated
samples are faithful descriptions of the collected data. Although the
simulation programs have been extensively tuned and validated over
years of use,  skepticism remains  about their ability to
accurately describe the correlations in a high-dimensional space,
leading to reasonable concerns about whether a network's learned
solution relies on a well-modeled physical effect or an overlooked
weak point.  This concern applies to shallow networks as well, but
is even more severe when working with higher-dimensional
lower-level data.

One means to assuage these concerns is a classic piece of
experimental scientific strategy: validation using adjacent
control regions. The primary concern is whether the correlations
among the input features are well modeled, which can be verified through a comparison of the network function evaluation in real and simulated
data. Adjacent control regions can be employed to keep the data that are most sensitive to the
hypothetical signal blind. 
This technique is often applied in the case of single-dimensional
data analysis, and is essentially a generalization of the
sideband approach.  If the input feature correlations that the network function relies on
are poorly modeled, the distributions of
the real and simulated data will disagree.  This provides some
validation of the network function, but no insight into its
structure.

An alternative approach relaxes the reliance on simulated samples
by using real data in the training step and avoids the need for
training labels by using {weakly supervised
learning}~\cite{Dery:2017fap,Metodiev:2017vrx}. In one approach,
one needs to know only the proportion of labels in different
subsets of data~\cite{Dery:2017fap}, for instance, samples of
events with known proportions of quark and gluon jets. One approach, known as classification without labels
(CWoLa), requires only that the different samples of events have
different label proportions even if they are
unknown~\cite{Metodiev:2017vrx}.

In other cases, the weak points of the modeling are well known to
physicists, who would prefer a learned solution that avoids
detailed reliance on these features. For example, most simulated
samples of collisions that result in jets use either {\sc
Pythia}~\cite{Sjostrand:2006za} or {\sc Herwig}~\cite{Bahr:2008pv}
to model the parton shower, but these two heuristic approaches can
make significantly different predictions about the jet
images~\cite{Barnard:2016qma}.  One way to address this issue is
to explicitly parameterize the network in the space of the unknown
nuisance parameter~\cite{Cranmer:2015bka,Baldi:2016fzo} so that
the dependence can be studied or constrained in data.  Another is
to attempt to explicitly reduce the dependence of the network on
aspects that are sensitive to the underlying
uncertainties~\cite{Louppe:2016ylz}, making the resulting network
less sensitive to these uncertainties.  This can be regarded as a
constrained optimization problem; for example, one might seek an
optimal combination of jet substructure tagging variables that
does not distort the smooth background~\cite{Shimmin:2017mfk}.

\subsection{Software}

The growing complexity of neural networks, in terms of both the
architecture and the raw computational power required for
training, might seem overwhelming to the pragmatic high-energy
physicist. Worse, the summary above is merely a brief review of
state-of-the-art deep learning, and given the rapid pace at which
the field evolves, techniques may change in the near
future.  Fortunately, a number of software packages~\cite{chollet2015keras,tensorflow,theano}
already provide efficient automatic gradient computation and
interfaces to hardware such as GPUs. These are well supported outside HEP. Thanks to these tools, the role of the physicist is reduced
to choosing an appropriate problem, data representation, architecture, and training strategy.

The
software landscape continues to evolve rapidly compared with
typical timescales for software in collider physics. This contrast
presents a challenge for any experiment using deep learning:
By choosing one of the dozen currently available frameworks, the
experiment risks being marooned with an unsupported and bloated
dependency when the deep learning industry moves on. Fortunately,
while projections into the future of a specific deep learning
package, or even a particular architecture, would be premature,
the underlying representation of a network as a stack of
differentiable tensor operations has proven quite robust.

As deep learning matures, the language and specifications become
more precise. Several  ongoing projects with significant commercial
backing~(\citen{onnx}; for a list of useful tools and references bridging the gap between
collider physics and machine learning, see the following
repository: https://github.com/iml-wg/HEP-ML-Resources) seek to formalize these specifications
further. Such formal specifications allow a factorization between
the training phase---which can depend on specific hardware and a
myriad of software packages---and the application or inference
phase. Training a neural network requires millions of iterations,
whereas inference requires only  a single pass per classified
pattern. As a result, computational demands at the inference stage
are mild. The factorization enables  inference-only
implementations~\cite{lwtnn} or autogenerated inference functions
to be the primary vehicle for incorporation into the trigger and
reconstruction software of the LHC experiments.

In contrast, as machine learning is incorporated into high-level data analysis, there is a benefit to having closer integration of modern machine learning frameworks and
statistical analysis software. Various tools are beginning to
blend deep learning, probabilistic modeling, and statistical
inference~\cite{carl,  edward,tran2016edward, pyroweb}.

\section{PROSPECTS}

In the past few years, advances in machine learning have enabled
the development of tools that have the power to transform the
nature of data analysis in HEP. 

 Expected
increases in computational power and further advances in
training strategies can be reasonably expected to extend  the
power of these tools.  But important questions remain regarding how to best apply this power. Should physicists aim for
end-to-end learning, giving the network the data at the
lowest level and highest dimensionality that it can effectively
process and ask it to solve the entire problem at once?  Or is it
more sensible to maintain outlines of the existing structures, but
replace engineered solutions based on domain knowledge with
learned solutions? A middle road, inspired by the hierarchical
nature of the LHC data and the success of highly structured
networks, suggests building end-to-end tools whose internal
structure reflects the outlines of existing analysis pipelines
(Figure~\ref{fig:embedding}). Planning for such future efforts
is already under way~\cite{Elmer:2017rej, Alves:2017she}.

Beyond the application of deep learning to the problems described
in this review, related research in the fields of statistical and machine
learning offer promising solutions to the challenges of data
analysis in particle physics. For example, the problem of modeling
smooth background distributions from observed data has long been
treated using ad hoc parametric functions; techniques from the
study of Gaussian processes have recently been shown to provide a
powerful and promising alternative~\cite{Frate:2017mai}.  Another
area with significant untapped potential with relevance to collider
physics is that of anomaly detection; recent
techniques~\cite{DBLP:journals/corr/LuoS17a} have improved the
data compression and anomaly detecting speed such that
applications to use these techniques to search for anomalous
signatures in the LHC data set may be practical.  Recently,
Bayesian optimization has been used for more efficient tuning
of the simulation programs~\cite{Ilten:2016csi}, and adversarial
training of GANs has been extended to the tuning of nondifferentiable
simulators~\cite{avo}. Finally, several groups have used machine
learning to grapple with the interpretation of results in
high-dimensional  parameter spaces for theories such as
supersymmetry~(\citen{Bertone:2016mdy,Caron:2016hib}; also see https://indico.cern.ch/event/632141/). \

One of the most profound developments in machine learning is the ability to model high-dimensional distributions from large samples of data. The ability to estimate high-dimensional probability densities or density ratios enables probabilistic inference in situations that were previously intractable. The key machine learning developments here~\cite{larochelle2011neural,2015arXiv150505770J,DBLP:journals/corr/DinhKB14,Cranmer:2015bka,kingma2016improving,Cranmer:2016lzt, papamakarios2017masked} allow the tuning of the classification tool for the particular problem at hand, opening the door to deeper levels of optimization and potentially more powerful analyses.   

Further work in this area includes efforts to modify the simulation tools for improved sampling of the high dimensional space\cite{le2016inference,ProbProg}. An exciting direction of this research is to automatically discover what sequence of events in the simulation of a background process leads to  rare events being misclassified as signal. 

Deep learning has already influenced data analysis at the LHC and sparked a new wave of collaboration between
the machine learning and particle physics communities, which is progressing
at a rapid pace. While it is difficult to predict the ultimate
impact these developments will have, we anticipate that new
applications will be found, motivating new strategies for
analysis of the LHC data and yielding deeper insights into fundamental questions in particle physics.

\section*{DISCLOSURE STATEMENT}

The authors are not aware of any affiliations, memberships,
funding, or financial holdings that might be perceived as
affecting the objectivity of this review.

\section*{ACKNOWLEDGMENTS}
The authors thank Ben Nachman for helpful comments.  DW and DG are supported by the Office of Science at the Department of Energy. KC is supported by the National Science Foundation (ACI-1450310 and PHY-1505463) and by the Moore-Sloan Data Science Environ- ment at at NYU.

\end{document}